\documentclass[aps,pra,twocolumn,showpacs,superscriptaddress]{revtex4-1} 

\usepackage{natbib} 
\usepackage[normalem]{ulem}
\usepackage{graphicx}
\usepackage{amssymb}
\usepackage{amsmath}
\usepackage{bm}
\usepackage{bbm}
\usepackage[caption=false,labelformat=simple]{subfig}
\usepackage{hyperref}

\usepackage{natbib}
\usepackage{color, soul}

\usepackage{graphicx}  % needed for figures
\usepackage{dcolumn}   % needed for some tables
\usepackage{bm}        % for math
\usepackage{amssymb}   % for math

\graphicspath{{./Figures/}}
\begin{document}

\widetext

\title{Rydberg optical Feshbach resonances in cold gases}
\author{N\'ora S\'andor} 
\affiliation{icFRC, IPCMS (UMR 7504) and ISIS (UMR 7006), University of Strasbourg and CNRS, 67000 Strasbourg, France}
\author{Rosario Gonz\'alez-F\'erez} 
\affiliation{Instituto Carlos I de F\'isica Te\'orica y Computacional and Departamento de F\'i­sica At\'omica, Molecular y Nuclear,
Universidad de Granada, 18071 Granada, Spain}
\author{Paul S. Julienne} 
\affiliation{Joint Quantum Institute, University of Maryland and NIST, College Park, Maryland 20742, USA}
\author{Guido Pupillo} 
\affiliation{icFRC, IPCMS (UMR 7504) and ISIS (UMR 7006), University of Strasbourg and CNRS, 67000 Strasbourg, France}

\date{\today}
\begin{abstract}
We propose a novel scheme to efficiently tune the scattering length of two colliding ground-state atoms by off-resonantly coupling the  scattering-state to an excited Rydberg-molecular state using laser light. For the $s$-wave scattering of two colliding  ${^{87}}\mathrm{Rb}$ atoms, we demonstrate that the effective optical length and pole strength of this Rydberg optical Feshbach resonance can be tuned over several orders of magnitude, while incoherent processes and losses are minimised. Given the ubiquity of Rydberg molecular states, this
  technique should be generally applicable to homonuclear atomic pairs as well as to atomic mixtures with $s$-wave (or even $p$-wave) scattering. 
\end{abstract}

\pacs{32.80.Ee,32.80.Qk,32.80.Rm}

\maketitle
\section{Introduction}
Magnetic Feshbach Resonances~\cite{Chin_2010} allow for tuning the strength of interactions in cold gases when the given atomic or molecular species possesses electronic ground states that are sensitive to magnetic fields. Magnetic Feshbach Resonances have been key tools for a number of experimental breakthroughs, including the realisation of strongly-correlated many-body quantum systems~\cite{Bloch_2008}, exotic few-body states~\cite{Braaten_2006,Greene_2010,Ferlaino_2011,Serwane2011,Pollack2009,Zaccanti_2009,DIncao_2013}, the production of cold molecules~\cite{Jin_2012,Kohler_2006, Ni_2008,Danzl_2008} and molecular BECs~\cite{Zwierlein2003, Greiner2003, Jochim2003}. For systems where the ground states are not magnetically sensitive, optical Feshbach resonances (OFR) have been proposed as an alternative to modify the scattering length, where
the resonance is created via laser-coupling~\cite{Bohn_1999, Theis_2004, Ciurylo_2005, Enomoto_2008, Yan_2013, Nicholson_2015}. This is a promising tool for, e.g., alkaline-earth-type atoms, where excited states can have exceedingly small line widths. %In these systems, OFRs have been already used to {\color{red}thermalise cold gases.}
For a generic atomic species, however, the utility of this technique is invariably limited by the finite lifetime of the excited molecular states. 

In this work, we propose and demonstrate that the scattering length of two colliding ground-state atoms can be efficiently tuned by coupling the two-atom ground-state to an excited Rydberg-molecular state using off-resonant laser light.
The ultralong-range Rydberg molecules~\cite{Greene_2000,Hamilton_2002, Bendkowsky_2009, Bendkowsky_2010,Anderson_2014,Niederprum_2016,Niederprum_2016a,Bottcher_2016} constitute a special group of molecules, 
where  molecular bonding is provided by the scattering between a Rydberg-electron and a ground-state atom. Coupling a Rydberg molecular state to the two-particle 
scattering state via laser light results in a Rydberg OFR.
 Key parameters for any OFR are the optical length and pole strength~\cite{Nicholson_2015} 
\begin{subequations} \begin{align}
\ell_{\rm opt}(\epsilon)=\frac{\Gamma_{\rm stim}(\epsilon)}{\gamma_m\sqrt{ 2\mu\epsilon}}\label{eqn:opt_length}\\ 
  %, characterising the photo-association line strength factor, 
 s_{\rm res}(\epsilon)=\frac{\ell_{\rm opt}(\epsilon)\gamma_{\rm m}}{\bar a \bar E},\label{eqn:pole_strength}\end{align}\end{subequations}  where $\Gamma_{\rm stim}(\epsilon)$ and $\gamma_m$ are 
the stimulated and spontaneous emission rates of the molecular state, respectively, $\bar{a}$ is the mean scattering length of the ground state van der Waals potential with a corresponding energy 
$\bar{E}=1/(2\mu \bar{a}^2)$~\cite{Gribakin_1993},  $\epsilon$ is the collision energy, i.e.,  
$k(\epsilon)=\sqrt{ 2\mu\epsilon}$ the relative momentum, and $\mu$ is the reduced mass~\cite{units}. 
Useful resonances are those for which both $\ell_{\rm opt}(\epsilon)/\bar{a}$ and $s_{\rm res}(\epsilon)$ are (much) larger than unity, which implies that the ratio of coherent to incoherent processes in the light coupling is large and that the resonance is broad and tuneable, respectively. Using an analytical model as well as numerically exact coupled channel calculations for $^{87}$Rb atoms, we demonstrate that the effective optical length and pole strength of this Rydberg optical Feshbach resonance can be tuned over several orders of magnitude. As a result, the real part of the scattering length can be tuned over a useful range of parameters, while incoherent processes are minimized over timescales as large as tens of milliseconds due to the comparatively small line widths of Rydberg states~\cite{Butscher_2011}. 
Varying the Rydberg principal quantum number, within a range that is experimentally accessible,
allows one to obtain results comparable to, or even better than, those obtained 
with traditional OFR for Sr and Yb atoms, for a broad range of atomic species or mixtures. 
%The  requisite on the atomic species is  a negative scattering length  for the  electron ground-state atom collision for the Rydberg molecular state to exist~\cite{Greene_2000}. 
For several alkali atoms, %for which the scattering length of the Rydberg-electron and ground state atoms is known, 
we provide examples of scaling of parameters within experimentally useful ranges. We note that the generality of the present 
technique can be extended to other situations, including, e.g., non bi-alkali mixtures  or manipulating $p$-wave scattering 
by optical means.

The paper is organized as follows: In~\autoref{sec:ROFER} we first describe the proposed technique for Rydberg OFRs, formulated as a three-channel scattering problem. \autoref{sec:RM} presents the Hamiltonian used to derive the Rydberg molecular states. ~\autoref{sec:scat} presents two complementary approaches to solve the scattering problem, based on numerical coupled channel calculations as well as on an approximate analytical estimate of the scattering length in the single-resonance approximation, respectively. The results for the real and imaginary parts of the scattering length for Rydberg OFR with Rubidium atoms for a fixed scattering energy, as well as for finite temperatures with Rb atoms and in mixtures of K and Cs atoms, are presented in~\autoref{sec:Res}, together with calculations of the thermal averaged $\ell_{\rm opt}^T$ and $s_{\rm res}^T$. Sec.~\autoref{sec:Sum} provides a summary and outlook of this work. 

\section{Interaction scheme for Rydberg Optical Feshbach Resonance\label{sec:ROFER}}

\begin{figure}[htp]

\subfloat[\label{Fig:1a}]{%
  \includegraphics[clip,width=\columnwidth]{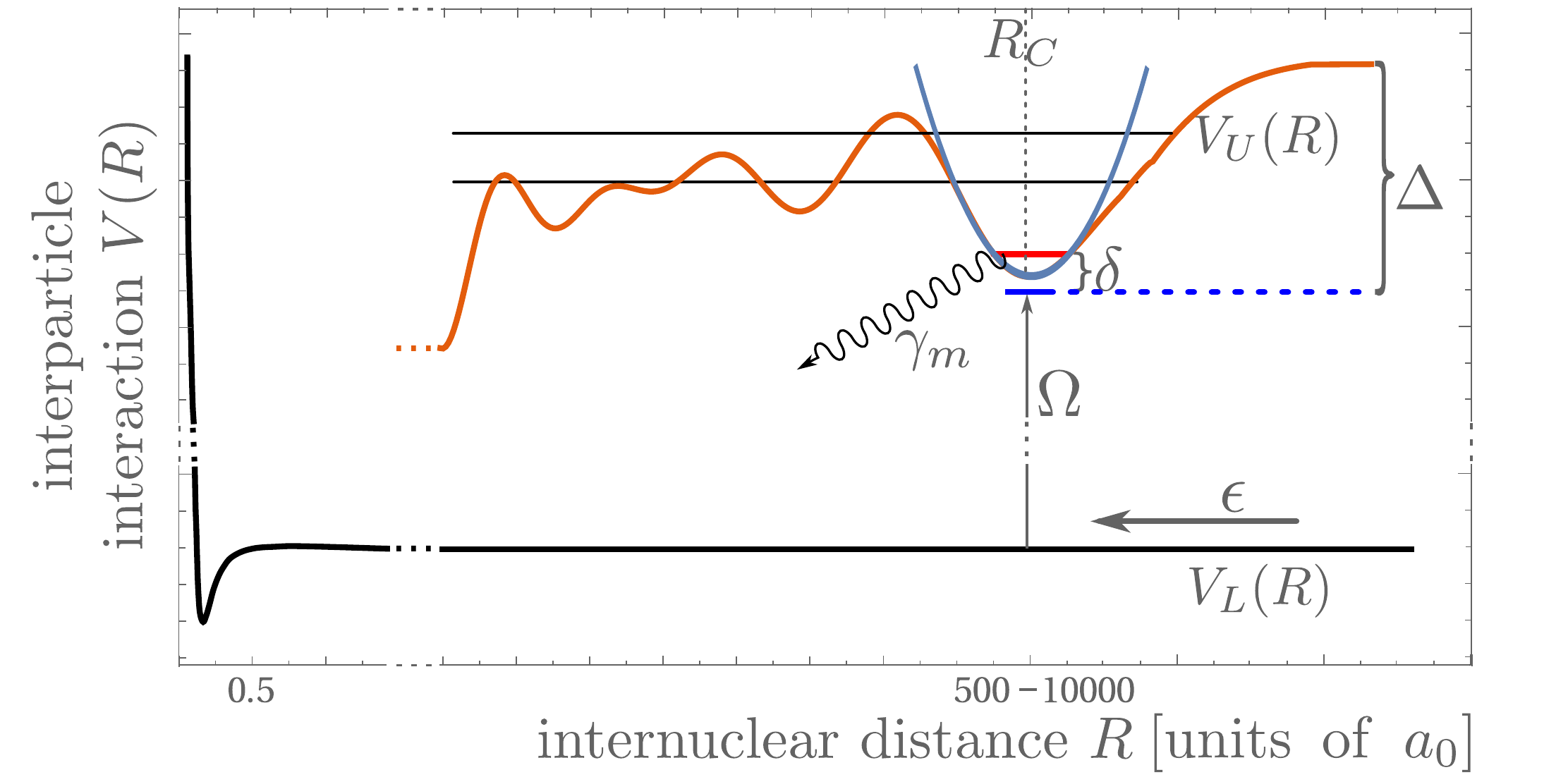}%
}

\subfloat[\label{Fig:1b}]{%
  \includegraphics[clip,width=\columnwidth]{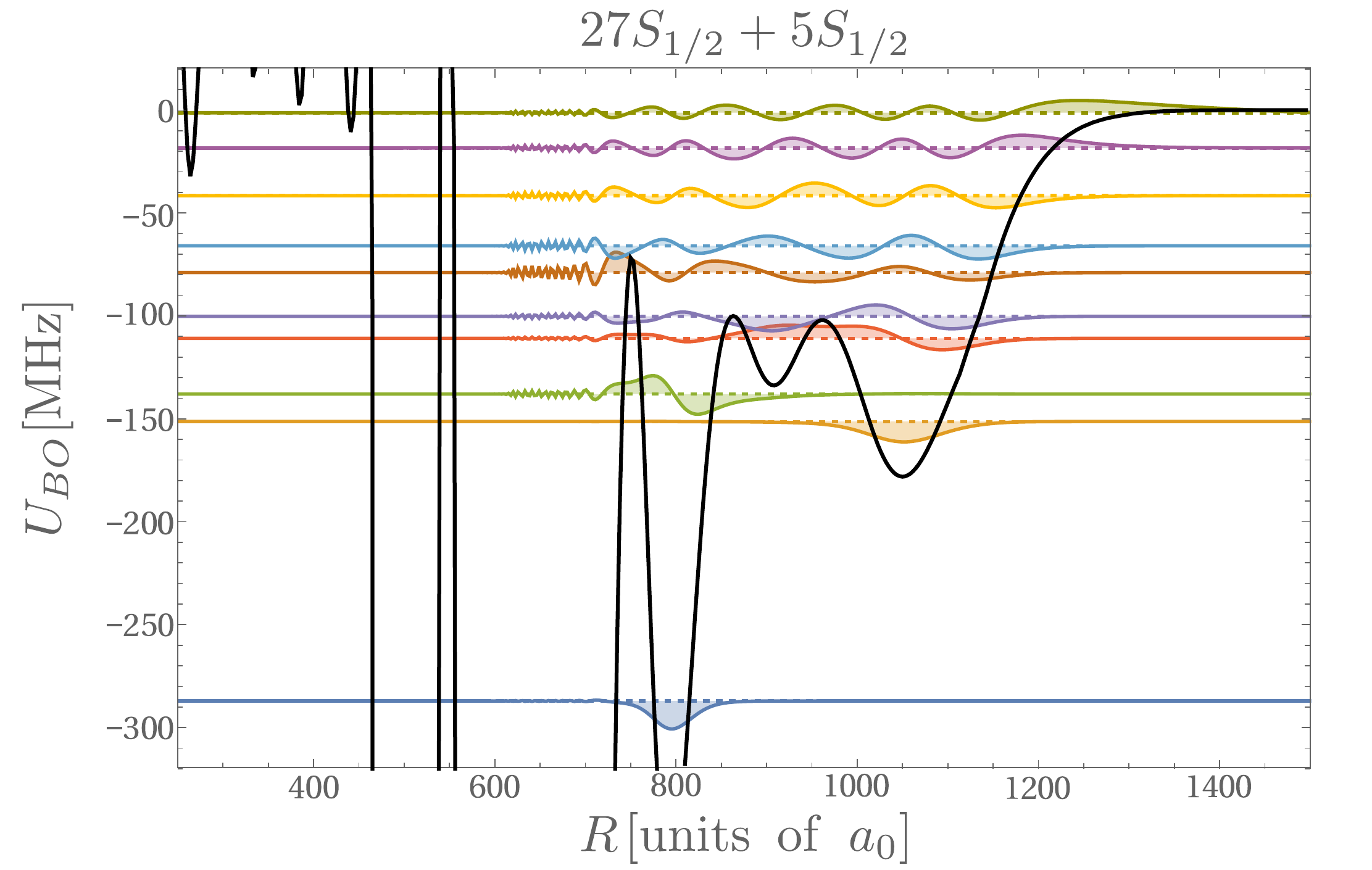}%
}

\caption{\label{Fig:1} (a): Sketch of a Rydberg optical Feshbach resonance: Two ground state atoms colliding with relative energy $\epsilon$ are off-resonantly coupled by laser light (vertical arrow) to an eigenstate of an excited Rydberg molecular potential with spontaneous emission rate $\gamma_{\rm m}$ (wiggling arrow). Here, $V_L(R)$ and $V_U(R)$ are the ground and Rydberg Born-Oppenheimer potentials, respectively\cite{units}. $\Omega$ is the laser Rabi frequency, $\delta$ the detuning from molecular resonance and $\Delta$ the detuning from the asymptotic atomic Rydberg state energy, and $R_C$ the Condon point.  
(b): Lowest-lying 
Rydberg molecular Born-Oppenheimer potential\cite{units} (continuous black line) %computed from Eq.~\eqref{Eqn:Hamiltonian} for 
of the 
%$\left|ns,5s\right\rangle$ $\mathrm{Rb}_2$
Rb($27S_{1/2}$)Rb($5S_{1/2}, F=1$)  $M_K=1/2$ dimer, together with the computed eigenstates of the potential (see text).}
\end{figure}

In the proposed scheme, sketched in~\autoref{Fig:1a},  the collision of two atoms in the 
presence of the light field is treated as an effective three-channel scattering problem. 
The entrance channel is a pair of atoms  in a $s$-wave scattering state of the ground-state 
electronic potential $V_L(R)$ with relative energy $\epsilon$. 
Here, we assume that one atom is always $^{87}$Rb(5$S_{1/2}$)~\cite{Strauss_2010}, and the second one is 
another alkali metal, such as ${^{87}}$Rb(5$S_{1/2}$), $^{41}$K(4$S_{1/2}$)~\cite{Pashov_2007} or 
${^{133}}$Cs(6$S_{1/2}$)~\cite{Takekoshi_2012}.
The excited channel is a bound state of  the  %${^3}\Sigma$ 
adiabatic potential $V_U(R)$  of the  Rydberg molecule~\cite{Greene_2000} formed by an  excited rubidium, Rb($nS_{1/2}$), 
and a ground state atom, which are 
bound due to the scattering of the Rydberg-electron with the ground-state perturber.
The third channel represents the product generated by spontaneous decay. In the calculations, we cautiously estimate the 
molecular decay rate $\gamma_m$ to be twice the spontaneous emission rate of the atomic Rydberg state~\cite{Beterov_2009, Branden_2010, lifetime}. 
In~\autoref{Fig:1a}, the vertical arrow represents the laser light, which couples off-resonantly the scattering state and the excited bound state 
with (two-photon) Rabi frequency 
$\Omega$ and red detunings $\delta$ and $\Delta$ from the chosen molecular resonance and the asymptotic atomic Rydberg 
energy, respectively, with $\Omega \ll \delta$ and $ \gamma_m \ll \delta$.

% 
% \begin{figure}[!h]
%  \includegraphics[width=0.5\textwidth]{fig1a}
%  \includegraphics[width=0.5\textwidth]{fig1b} 
% \caption{\label{Fig:1} (a): Sketch of a Rydberg optical Feshbach resonance: Two ground state atoms colliding with relative energy $\epsilon$ are off-resonantly coupled by laser light (vertical arrow) to an eigenstate of an excited Rydberg molecular potential with spontaneous emission rate $\gamma_{\rm m}$ (wiggling arrow). Here, $V_L(R)$ and $V_U(R)$ are the ground and Rydberg Born-Oppenheimer potentials, respectively\cite{units}. $\Omega$ is the laser Rabi frequency, $\delta$ the detuning from molecular resonance and $\Delta$ the detuning from the asymptotic atomic Rydberg state energy, and $R_C$ the Condon point.  
% (b): Lowest-lying 
% Rydberg molecular Born-Oppenheimer potential\cite{units} (continuous black line) %computed from Eq.~\eqref{Eqn:Hamiltonian} for 
% of the 
% %$\left|ns,5s\right\rangle$ $\mathrm{Rb}_2$
% Rb($27S_{1/2}$)Rb($5S_{1/2}, F=1$)  $M_K=1/2$ dimer, together with the computed eigenstates of the potential (see text).}
% \end{figure}

The off-resonant coupling takes place at the minimum $R_C$ of the ultralong-range potential $V_U(R)$, which depends on the principal quantum number of the Rydberg state (see the inset of \autoref{Fig:2}) and can be found at distances $R_C\sim10^3-10^4a_0$, much larger 
than the typical ones for usual OFRs with $R_C\sim 10-10^2 a_0$, with $a_0$ being the Bohr radius. 

We assume that the gas is dilute  $d\gtrsim R_C$,  with $d$ being the average interparticle distance, so that the scattering 
problem can be described by a mean-field type energy-dependent scattering length $\alpha(\epsilon)$. This diluteness assumption is satisfied in cold and ultracold gases, with the maximum density of  $\mathfrak{n} \lesssim 10^{12}\mathrm{cm}^{-3}-10^{15}\mathrm{cm}^{-3}$\cite{units}, respectively. Here we estimated the principal quantum number-dependent maximum density as $\mathfrak{n}=(2\times R_C)^{-3}$.
%$\mathfrak{n} \sim 8\times 10^{13}-8\times 10^{15}\mathrm{cm}^{-3}$. 

%While satisfying the assumption of diluteness, the scheme is applicable in gases having the density of $\mathfrak{n} \sim 8\times 10^{13}-8\times 10^{15}\mathrm{cm}^{-1}$. This density range corresponds to cold, as well as ultracold atomic ensembles.

We further assume that the laser light mainly couples one selected bound state of $V_U(R)$, while the other bound states  of $V_U(R)$ [horizontal 
lines~\autoref{Fig:1a} and ~\autoref{Fig:1b}], as well as three- and higher-body resonances~\cite{Bendkowsky_2010}, are far detuned from 
the chosen bound state of the Rydberg molecular potential, a situation that is readily  obtained, e.g., 
for red detuning. Our scheme is useful when the real part of $\alpha$ is much larger than the imaginary part. In the 
following, we provide detailed calculations that prove that this regime can be readily attained for a variety of situations of 
experimental interest.

\subsection{Diatomic Rydberg molecular state}\label{sec:RM}

For the diatomic  Rydberg molecule, we assume that the ground-state atom and the Rydberg core can be treated as point 
particles. We work on  the low-energy regime, in which the interaction between the Rydberg 
electron and the neutral atom is well  approximated by
a Fermi pseudopotential~\cite{Fermi,Omont_1977}. In addition, we include the hyperfine interaction 
and the fine structure for the ground-state and Rydberg atoms~\cite{Anderson_2014,Markson2016}, respectively. Note that we are neglecting the  hyperfine interaction of the Rydberg atom because, since its energy decreases as $n^{-3}$\cite{Tauschinsky_2013}, it is negligible compared to the other terms in the Hamiltonian.
In the Born-Oppenheimer approximation,  the Hamiltonian of this diatomic  Rydberg molecule reads
\begin{eqnarray}
\label{Eqn:Hamiltonian}
	    \hat{H}(\mathbf{r},Z)&=&\hat{H}_0+ 
	    \sum_{i=S,T}2\pi A_s^i(\kappa)\delta^3(\mathbf{r}-Z\hat{\mathbf{z}})\hat{\mathbb{I}}_i
	     \\
	    %+2\pi A_s^S(k)\delta^3(\mathbf{r}-Z\hat{\mathbf{z}})\hat{\mathbb{I}}_S
	    &+& \sum_{i=S,T}6\pi A_p^i(\kappa)\delta^3(\mathbf{r}-Z\hat{\mathbf{z}})\overleftarrow{\nabla}\cdot\overrightarrow{\nabla}\hat{\mathbb{I}}_i+ A_{\rm hf}\color{black} \hat{\mathbf{S}}_g\cdot \hat{\mathbf{I}}_g \,, \nonumber	   
\end{eqnarray}
where, 
$\hat H_0$ is the single electron Hamiltonian describing the Rydberg-atom, and $\kappa$ 
is the momentum of the Rydberg-electron, which  in the semiclassical approximation is given by $\kappa=\sqrt{2/R-1/n^{\ast 2}}$ and 
$n^{\ast}=n-\delta_\ell$, with $n$ being the principal quantum number and $\delta_\ell$ the quantum defect.  
The second and third terms are the $s$- and $p$-wave Fermi pseudopotential describing the interaction
between the Rydberg electron and the ground-state perturber,  $A_{s,p}^{T,S}(\kappa)$ are the non-relativistic energy-dependent $s$- 
and $p$-wave, triplet and singlet,  scattering lengths between the Rydberg electron and the ground-state atom~\cite{Fabrikant_1986}, and 
$\hat{\mathbb{I}}_T=\hat{\mathbf{S}}_r\cdot\hat{\mathbf{S}}_g+3/4$ and $\hat{\mathbb{I}}_S=\hat{\mathbb{I}}-\hat{\mathbb{I}}_T$ 
are projectors on  the triplet and singlet scattering channels,  respectively, with 
$\hat{\mathbf{S}}_r$ and $\hat{\mathbf{S}}_g$  the Rydberg-electron and ground state atom spins, respectively. 
The last term in the Hamiltonian~\eqref{Eqn:Hamiltonian} stands for the hyperfine interaction of the ground-state 
atom~\cite{Anderson_2014}, with $\hat{\mathbf{I}}_g$ being  the nuclear spin of the ground-state atom,  and $A_{\mathrm{hf}}$ the hyperfine coupling parameter. The total spin of the ground state atom is $\mathbf{F}=\mathbf{S}_g+\mathbf{I}_g$, with $\mathbf{S}_g$ and $\mathbf{I}_g$ being the electronic and nuclear spins respectively. Due to the azimuthal symmetry, the total magnetic quantum number is conserved $M_K=M_F+M_{S_r}$, with  $M_F$ and $M_{S_r}$ being the magnetic quantum numbers of the total spin of the ground state atom, $\mathbf{F}$, $M_F=M_{I_g}+M_{S_g}$, and the spin of the Rydberg atom $\mathbf{S}_r$, respectively.

In our calculations, we consider the Rydberg degenerate manifold Rb($(n-3),l\geq3$), and the energetically
closest neighboring Rydberg levels Rb($(n-2)d$), Rb($(n-1)p$), and Rb($ns$)~\cite{functions}.

\subsection{The scattering in the light-field}\label{sec:scat}

The Schr\"odinger equation of the scattering between the two ground-state atoms  reads
\begin{equation}
\label{eq:scattering}
	    \left[\frac{\partial^2}{\partial R}+2\mu(\epsilon\mathbb{I}-V(R)) \right]\Psi(R,\epsilon)=0\color{black}
\end{equation}
with $\mathbb{I}$ being the identity matrix and the potential matrix given by 
\begin{align}
V(R)=\begin{pmatrix}
     V_L(R)+V^{\infty}&\Omega\\
    \Omega&V_U(R)+V^{\infty}-\omega_{\mathrm{L}\color{black}}-i\frac{\gamma_{m}}{2}
     \end{pmatrix}, 
\end{align}
where $V_U(R)$ is the Born-Oppenheimer potential obtained from  
the diatomic Rydberg molecule Hamiltonian~\eqref{Eqn:Hamiltonian}, and  $V^{\infty}=-1/(2\Delta)+1/2\sqrt{\Delta^2+4\Omega^2}$ is the asymptotic light shift, with $\Delta$ being the detuning of the coupling laser having frequency $\omega_{\mathrm{L}\color{black}}$ from the atomic resonance, and $\Omega=\Omega_1\Omega_2/(2\delta)$ is the effective (real) Rabi-frequency of the two-photon coupling of the Rydberg-molecule\cite{Dudin2012}. Here $\Omega_1$ and $\Omega_2$ are the Rabi-frequencies which couple the ground-state and the Rydberg-state to the intermediate state, respectively, whereas $\delta$ is the detuning from the intermediate state. The experimental values of the effective two-photon Rabi frequency $\Omega$ for Rb atoms are usually in the range $\Omega/(2 \pi)\simeq0.1-10$ MHz. 

For the electronic ground-state potential, we  use the  Lennard-Jones form $V_L(R)=C_6[(\sigma/R)^6-1]/R^6-C_8/R^8$, 
where the  parameters $C_6$, $C_8$, and $\sigma$ are chosen to reproduce the ground-state scattering length of the 
considered species, see Ref.~\cite{values}.
The Schr\"odinger  equation~\eqref{eq:scattering} is solved using the Numerov propagator method  for
\begin{equation*}
\Psi(R,\epsilon)=\sum_{j\in\{1,2\}}{\frac{F_j(R,\epsilon)}{R}\left|j\right\rangle}
\end{equation*}
where $F_j(R,\epsilon)$ represents the amplitude of the wave-function on the basis function $\left|j\right\rangle$~\cite{Nicholson_2015}.

The numerical result of Eq.~\eqref{eq:scattering}   provides the $\alpha(\epsilon)$ scattering length between the two ground-state atoms; taking into account the energy dependence as well as the shifts due to off-resonant coupling of the laser field to all excited states in the Rydberg molecule potential $V_U(R)$.
As an example of $V_U(R)$, we present 
the lowest-lying  adiabatic potential of the Rydberg molecule Rb(27$S_{1/2}$)Rb(5$S_{1/2}, F=1$)
and total magnetic quantum number M$_K=1/2$ in \autoref{Fig:1b}, which has outer minima around $R_C\sim$800 and 1050$a_0$. 
The lowest-energy eigenvalues of these potential minima are well isolated from the rest of the spectrum, 
and, in our Rydberg OFR scheme, the laser is off-resonantly coupled to one of these bound states.

As a complementary approach to the coupled channel calculations, the influence of the laser that couples the colliding pair of atoms to the Rydberg bound state on the complex scattering length of these two ground-state atoms can also be analytically estimated within the single resonance approximation as~\cite{Nicholson_2015}
\begin{equation}
  \alpha(\epsilon)=\alpha_{\rm bg}(\epsilon)+\frac{\Gamma_{\rm stim}(\epsilon)\frac{[1+k(\epsilon)^2\alpha_{\rm bg}(\epsilon)^2]}{k(\epsilon)}}
  {\epsilon-\delta-k(\epsilon)\alpha_{\rm bg}(\epsilon)\frac{\Gamma_{\rm stim}(\epsilon)}{2}+i\frac{\gamma_{\rm m}}{2}}, \label{Eqn:scat}
\end{equation} %where $\alpha_{\rm bg}$ is the background scattering length. 
  with  $\alpha_{\rm bg}(\epsilon)$ being
 the energy-dependent, background scattering length of the two ground state atoms,
 $\Gamma_{\rm stim}(\epsilon)$ and $\gamma_m$
the stimulated and spontaneous emission rates of the molecular state, respectively,
and $k(\epsilon)=\sqrt{2\mu\epsilon}$. This approach allows us to analyze the general features of the scheme for a wide range of principal numbers of the Rb atom, and for different alkali atomic species such as ${^{87}}\mathrm{Rb}$, ${^{40}}\mathrm{K}$ and  ${^{133}}\mathrm{Cs}$.
 
The stimulated emission rate $\Gamma_{\rm stim}(\epsilon)$ coupling the scattering and excited bound
states has the usual form $\Gamma_{\rm stim}(\epsilon)=2\pi \Omega^2\left|F_{\epsilon}\right|^2$. Here,  $F_{\epsilon}=\int\psi_L(\vec{R},\epsilon) \psi_U(\vec{R})d\vec{R}$ is the Franck-Condon factor of the transition between the 
scattering state $\psi_L(\vec{R},\epsilon)$ and the bound one $ \psi_U(\vec{R})$. 
For the considered free-bound transition,  $F_{\epsilon}$ can be estimated analytically using a harmonic oscillator
approximation for the bound state as
\begin{align}
F_{\epsilon,n}=\sqrt[4]{\frac{8}{\pi\epsilon\omega(n)}}e^{\epsilon/\omega(n)}\sin\left[\sqrt{2\mu\epsilon}(R_C-\alpha_{\rm bg})\right],\label{eqn:FC01}
\end{align} 
with $\omega(n)$ being the frequency of the  harmonic potential fitted to the outermost minimum of the Rydberg 
electronic potential (see below).

The expression of the Frank-Condon factor shows that it depends not only on the energy but also on $R_C$ and on the ground state scattering length 
of the colliding atoms $\alpha_{\rm bg}(\epsilon)$, the zero-energy value of which is given in~\autoref{Tbl:perturber}. 
Since $R_C$ lies largely outside the molecular core region of the
two ground-state atoms, $\Gamma_{\rm stim}\sim \sin^2[k(\epsilon) (R_C-a_{bg})]$ instead of the usual 
dependence $\Gamma_{\rm stim}\sim k(\epsilon)$ for scattering at $\mu$K temperatures, similar to 
the behaviour observed in photoassociation of ultralong-range potentials~\cite{Jones_2006}. 
This implies that $\Gamma_{\rm stim}(\epsilon)$, and thus $\ell_{\rm opt}(\epsilon)$
and $s_{\rm res}(\epsilon)$, acquire a significant energy dependence (see Eq.~\eqref{Eqn:scat}).

\begin{table}[t]
\begin{tabular}{|c| c| c| c| c|}
 \hline
 perturber atom&$a_S^{p-e}\enspace[a_0]$&$\alpha_p[a_0^3]$&$\mu\enspace[a.u.]$&$\alpha_{bg}(\epsilon=0)\enspace[a_0]$\\
 \hline
 ${^{87}}\mathrm{Rb}$&-18.5&319&79295.6&100.4\\
 \hline
 ${^{40}}\mathrm{K}$&-15.4&303&57127.6&-186\\
 \hline
 ${^{133}}\mathrm{Cs}$&-21.7&402&95875.6&650\\
 \hline
 \end{tabular}\caption{\label{Tbl:perturber} 
Parameters controlling the strength of the Rydberg OFR induced by the coupling of  the scattering state to
a bound state of a  Rydberg molecular state of the  Rydberg molecules 
Rb($nS_{1/2}$)$X$, with $X$ being the ground-state of alkali perturber atoms ${^{87}}\mathrm{Rb}$, ${^{40}}\mathrm{K}$ and  ${^{133}}\mathrm{Cs}$, respectively.
$a_S^{p-e}$ is the zero-energy scattering length of the electron ground-state atom collision,
 $\alpha_p$ is the ground state polarizability,  $\mu$  is the reduced mass of the molecule,
 and $\alpha_{bg}(\epsilon=0)$ is zero-energy scattering length of the two ground-state atoms colliding.}
\end{table}

\begin{figure}[!hbt]
 \includegraphics[width=0.5\textwidth]{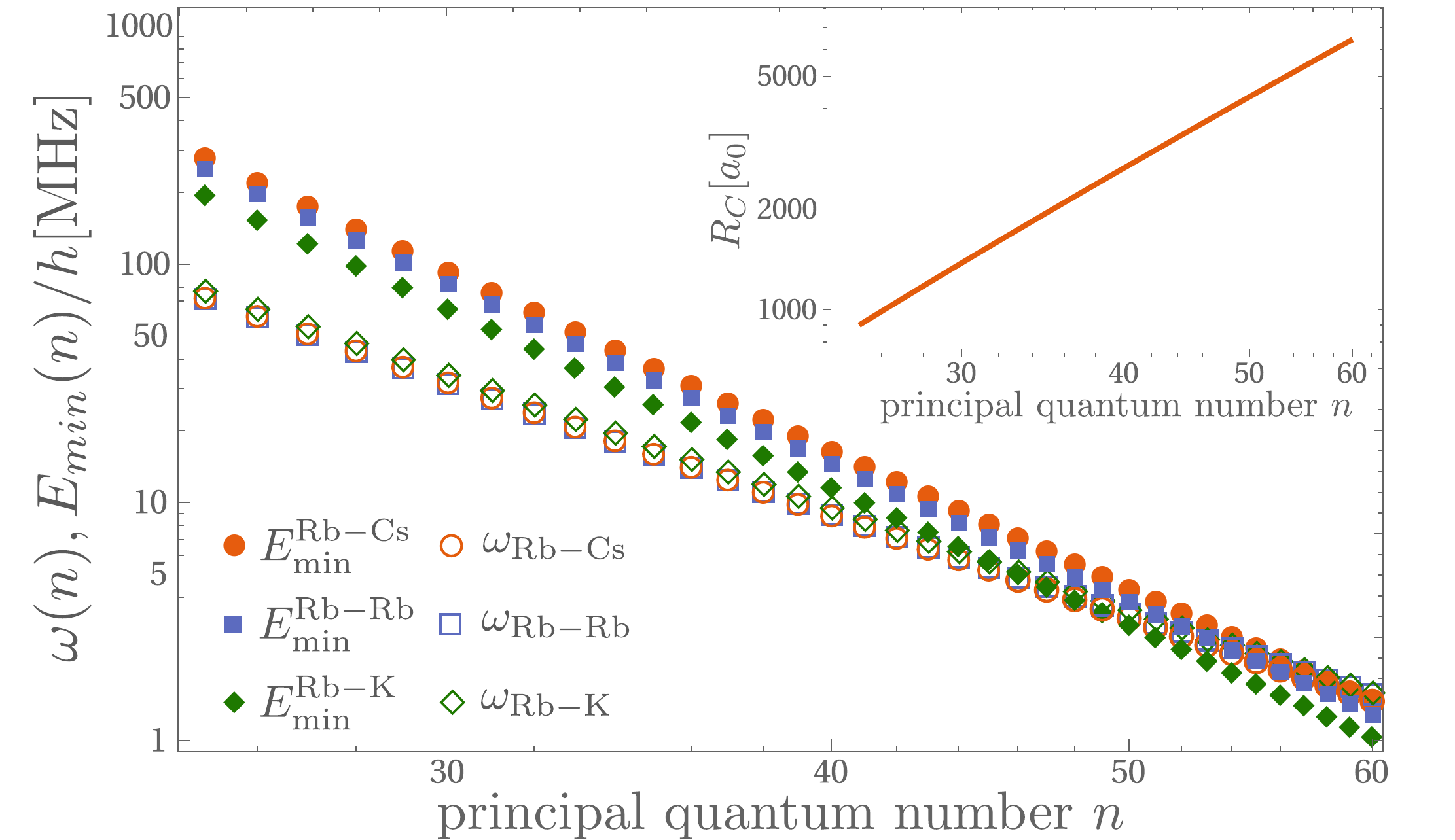}
 \caption{\label{Fig:2}
Parameters of the harmonic approximation of the outermost minimum of 
Rydberg-molecular state plotted in~\autoref{Fig:1b}  $\omega(n)$ is the frequency of the fitted harmonic potential, 
$|E_{min}(n)|$ is the corresponding depth of this potential, $R_C$ is the principal quantum number-dependent location of this minimum used 
in the Rydberg OFR\cite{units}.}
\end{figure}

For the Rydberg molecule Rb($nS_{1/2}$)$X$, with $X$ being the ground-state of alkali perturber atoms ${^{87}}\mathrm{Rb}$, ${^{40}}\mathrm{K}$ and  ${^{133}}\mathrm{Cs}$, 
 the frequency $\omega(n)$ and depth $E_{\rm min}(n)$ of the harmonic fit to the outermost minimum of the Rydberg 
 molecular potential, as sketched in~\autoref{Fig:1}, are presented as a function of the principal quantum number $n$  
 of the Rydberg excitation in~\autoref{Fig:2}.  For this harmonic fit, the  Born-Oppenheimer potential evolving from the Rydberg state Rb($nS_{1/2}$)
is approximated by the solution of the Schr\"odinger equation of  Hamiltonian~\eqref{Eqn:Hamiltonian} in the 
one dimensional Rydberg manifold $\mathrm{Rb}(ns)$. 
We are assuming that the contributions due to the  p-wave Fermi pseudo-potential,  
the singlet atom-electron scattering, and the neighbouring Rydberg-manifolds could be 
neglected~\cite{Bendkowsky_2009} in this approximation.
This analytic approximation reads  
$V_U(R)=2\pi A_s^T[\kappa] |\psi_{ns}(\vec{R})|^2$~\cite{Greene_2000}, where
the single s-wave scattering length between the electron and the ground state perturber is given by
$A_s^T[\kappa]=a_s^{p-e}+\pi\alpha_p\kappa/3$, with 
 $a_s^{p-e}$ being the zero-energy triplet scattering length between the electron and the ground state atom,
 $\alpha_p$  the ground state polarizability of the neutral perturber atom, 
 which are both given in~\autoref{Tbl:perturber},
  $\kappa$ the momentum of the Rydberg electron,
   and $\psi_{ns}(\vec{R})$ the wave-function of the  electron in the $ns$ Rydberg-state.
The features of this harmonic fit  only depend on the 
 scattering length between the  electron and ground state atom,
 which is of the same order of magnitude for the considered alkali atoms $\mathrm{Rb}$, $\mathrm{Cs}$ and 
 $\mathrm{K}$ (cf.~\autoref{Tbl:perturber}),  on  the wave function of the Rydberg electron of Rubidium,
 and on the Rydberg electron momentum $\kappa$.
 The frequencies $\omega(n)$ [entering Eq.~\eqref{eqn:FC01}] are generally found to be in the tens of MHz range, and 
 to decrease algebraically with $n$ as $\omega(n)\propto n^{-4}$ and $E_{\rm min}(n)\propto n^{-6}$. Thus,
based on the values of $\omega(n)$  and $E_{\rm min}(n)$ provided in~\autoref{Fig:2}, 
the  mixed-species Rydberg molecules formed by a $\mathrm{Rb}$ Rydberg atom   and $\mathrm{Cs}$ or  
$\mathrm{K}$ are expected to have similar features as those presented in~\autoref{Fig:1}.

In this work, we focus on Rydberg molecules formed from Rydberg states with  $n\lesssim 50$, as for larger $n$ { the energy-difference between the molecular state and the Rydberg atomic state is within the order of magnitude of the Rabi-frequency of the coupling laser, thus the single resonance approximation in Eq.~\eqref{Eqn:scat} is not applicable.}

\section{Results}\label{sec:Res}

\begin{figure}[htp]

\subfloat[\label{Fig:3a}]{%
  \includegraphics[clip,width=\columnwidth]{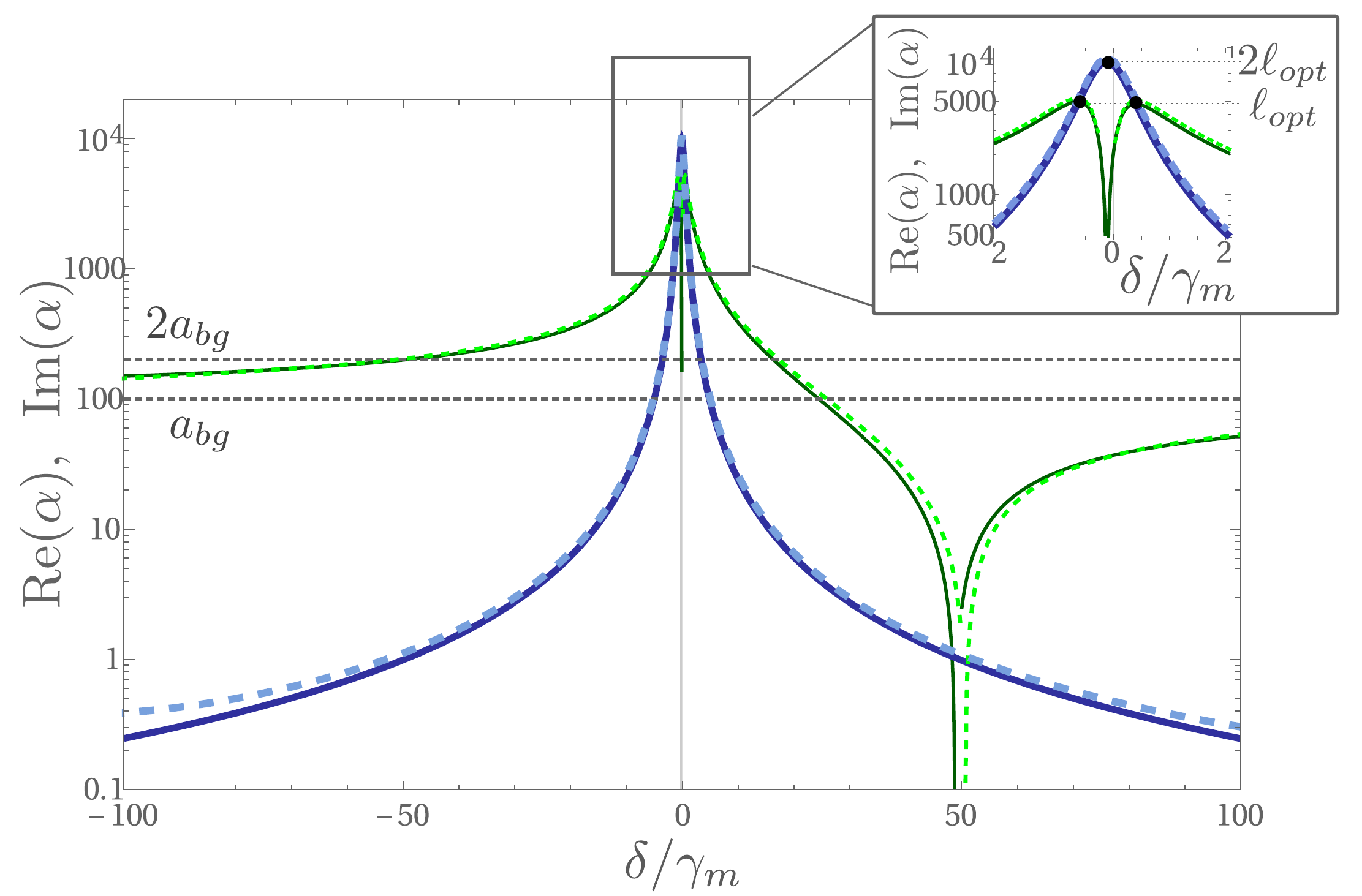}%
}

\subfloat[\label{Fig:3b}]{%
  \includegraphics[clip,width=\columnwidth]{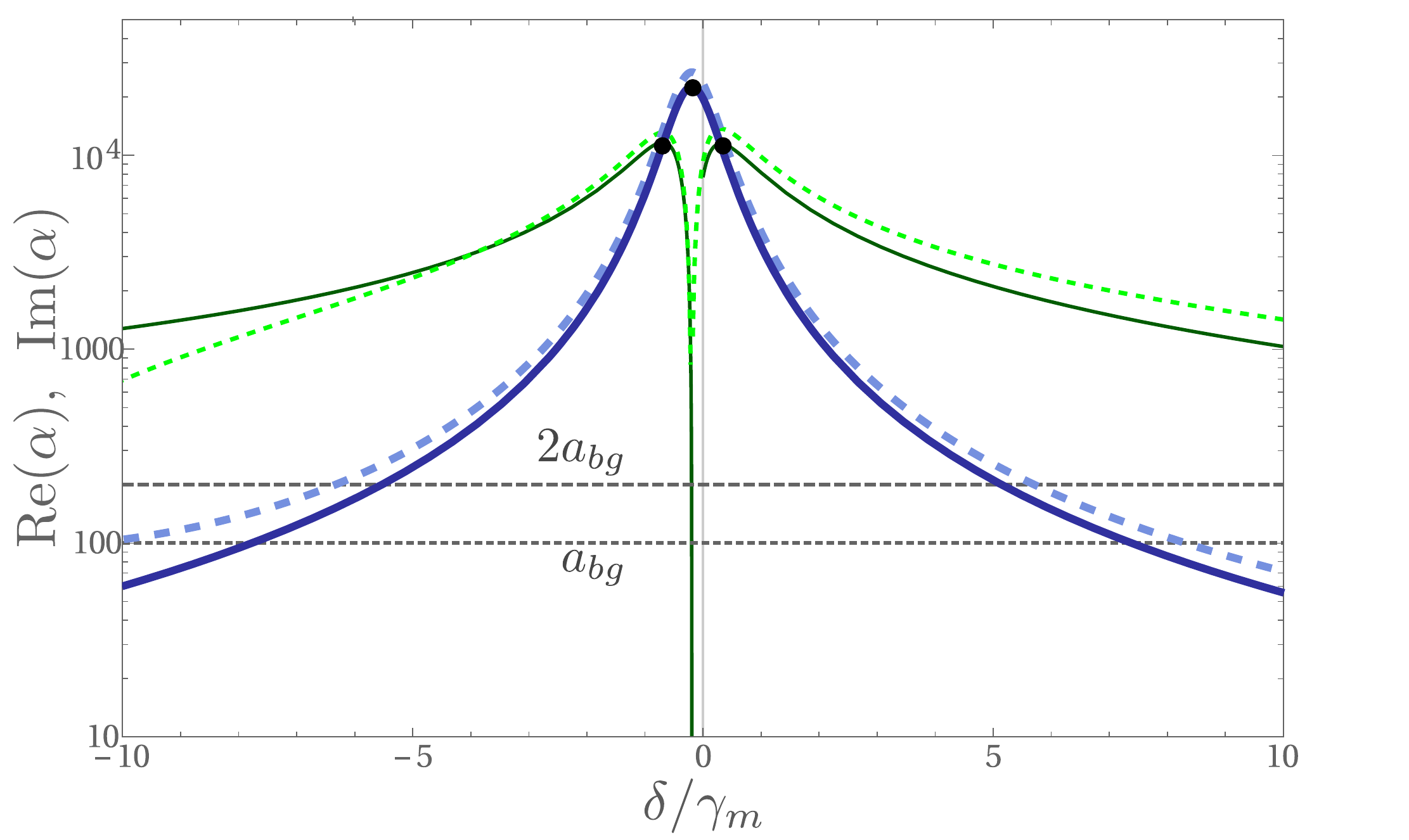}%
}

\caption{\label{Fig:3} 
 Analytical  prediction  Eq.~\eqref{Eqn:scat} (solid line)
 and coupled channel results  (dashed line) of the  real (green thin lines) and imaginary (blue tick lines) parts of the scattering length $\alpha(\epsilon)$ versus the detuning $\delta/\gamma_m$ from the lowest-lying molecular states of the dimers
(a) Rb($27S_{1/2}$)Rb($5S_{1/2}, F=1$)  and  (b) Rb($40S_{1/2}$)Rb($5S_{1/2}, F=1$) and 
$\enspace\Omega/(2\pi)=0.5\enspace\mathrm{MHz}$ and $0.2\enspace\mathrm{MHz}$ respectively.
The two ground-state Rb atoms collide  with relative energy $\epsilon=1\mu K\cdot k_B$, with $k_B$ being the Boltzmann-constant. 
The analytical (coupled channel) value of the optical length $\ell_{\rm opt}(\epsilon)$ is $\ell_{\rm opt}(\epsilon) \simeq 4900 a_0$ ($5270 a_0$) and 11500$a_0$ ($13400 a_0$) for panels (a) and (b), respectively. The linewidth of the molecular state is $\gamma_m/(2\pi) \simeq 144\mathrm{kHz}$ and  $\gamma_m/(2\pi)\simeq 54\mathrm{kHz}$ in panels (a) and (b), respectively.}
\end{figure}

\begin{figure}[htp]

\subfloat[\label{Fig:4a}]{%
  \includegraphics[clip,width=\columnwidth]{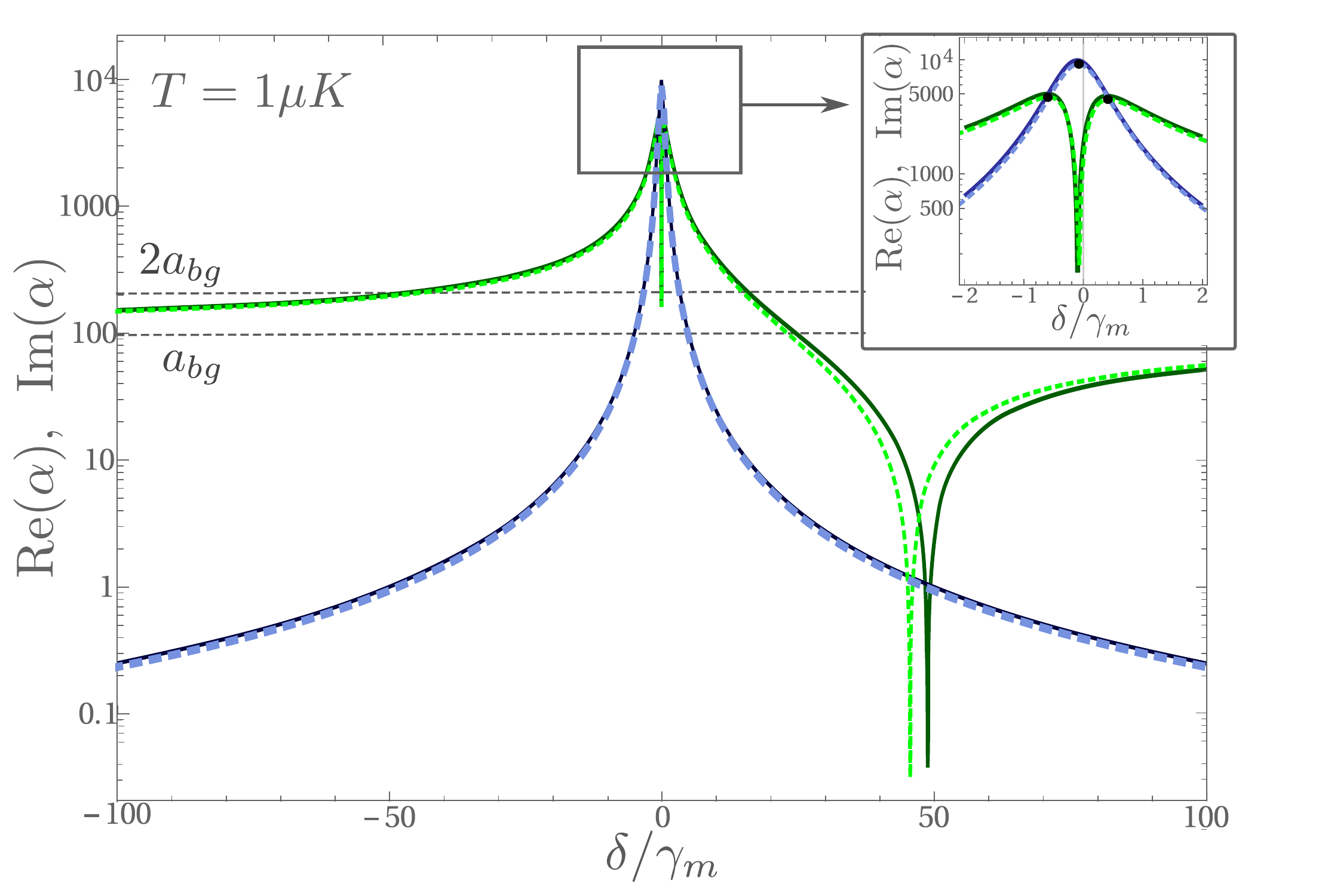}%
}

\subfloat[\label{Fig:4b}]{%
  \includegraphics[clip,width=\columnwidth]{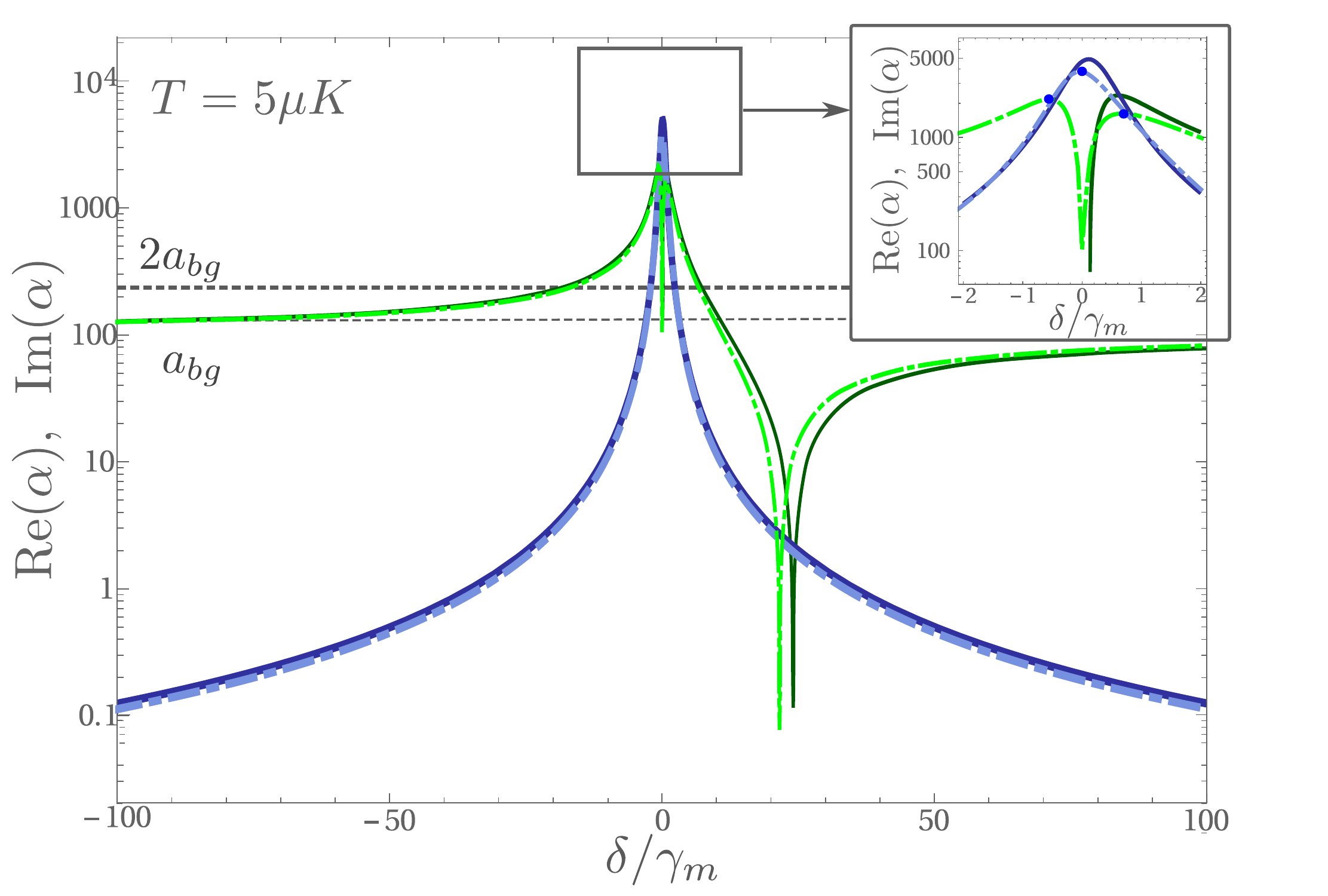}%
}

\caption{\label{Fig:4}Comparison between the thermal average of the scattering length $\bar \alpha^T$ (dashed lines), see definition in Eq.~\eqref{eqn:th_Av}, and the energy-dependent scattering length $\alpha(\epsilon)$
 (solid lines) for the molecular state of Rb($27S_{1/2}$)Rb$(5S_{1/2}$). The real and imaginary parts of  the 
 thermal average and the energy-dependent  scattering lengths are indicated by thin-green and thick-blue lines, respectively. The results   are obtained with the single resonance approximation~\eqref{Eqn:scat}, and calculated for
 the Rydberg excitation Rb($27S_{1/2}$),  temperatures $T=1 \mu K$ and $5 \mu K$ (dashed lines),  
 energies  $\epsilon=2\pi\cdot10^{-12}k_B~\hbar \mathrm{MHz}$ and $\epsilon=2\pi\cdot5\cdot10^{-12}k_B~\hbar \mathrm{MHz}$(solid lines).
  Insets: details of the real and imaginary parts of the scattering lengths around $\delta=0$. The  thermally averaged scattering length is smaller than the single energy scattering length and also show an asymmetric dependence on $\delta$ (in contrast with the symmetrical behavior of the single energy curves.)}
\end{figure}

Useful resonances are those for which the ratio of the real and imaginary  parts of $\alpha$,
$\mathrm{Re}(\alpha)$ and $\mathrm{Im}(\alpha)$, respectively, 
 is much larger than one, while $\mathrm{Re}(\alpha)$ differs significantly from the background scattering length $\alpha_{\rm bg}(\epsilon=0)$. 
 \autoref{Fig:3} shows that these conditions are fulfilled for a broad range of parameters with Rb atoms. 
For the electronic potential of 
Rb($27S_{1/2}$)Rb($5S_{1/2}, F=1$) and $\Omega/(2\pi)=0.5$MHz\cite{units}, the scattering length in~\autoref{Fig:3a} can 
be modified by as much as 100 percent by varying $\delta$ over a range of detunings as large as $\delta\sim10^2\gamma_{\rm m}$, 
implying reduced decoherence from spontaneous emission.

 For {\it red detunings}, we find that the ratio $\mathrm{Re}(\alpha)/\mathrm{Im}(\alpha)$ is large over a broad range of parameters. For instance, for $\delta\simeq 50 \gamma_m$ in case of the molecular state Rb($27S_{1/2}$)Rb($5S_{1/2}, F=1$), this would imply $\mathrm{Re}(\alpha)\simeq 2\alpha_{\rm bg}$ within a timescale of $(k_2 \mathfrak{n})^{-1} \gtrsim 5$ms, where the loss rate constant for ground state atoms is $k_2 = 8\pi (\hbar/\mu)\mathrm{Im}(\alpha)$\cite{Nicholson_2015} and we estimated the ground-state atomic density as $\mathfrak{n}\lesssim  10^{14} \mathrm{cm^{-3}}$\cite{units}.

In contrast, for {\it blue detunings}, the presence of higher-energy molecular resonances, i.e., higher excited vibrations inside  
the Rydberg potential, results in interferences with $\mathrm{Re}(\alpha)\simeq 0$,
i.e., around $\delta=50\gamma_m$ in \autoref{Fig:3}, a behaviour qualitatively similar to that of usual OFRs in the presence of several photo-association resonances in the excited state potential~\cite{Nicholson_2015}.

The results in~\autoref{Fig:3} show that the analytical approximation for $\alpha(\epsilon)$ (solid lines) is in 
excellent agreement with the numerical coupled channel calculations in all parameter regimes, which establishes the usefulness of 
Eq.~\eqref{Eqn:scat} for Rydberg OFRs; 
 see also $\ell_{\rm {opt}}(\epsilon)$ in the inset and caption of~\autoref{Fig:3}.
 We find that the terms proportional to $k \Gamma_{\rm stim}(\epsilon)$ in Eq.~\eqref{Eqn:scat} are crucial to ensure this agreement, which demonstrates the unusually large energy dependence of these resonances.
 
Good agreement between the analytical and coupled channel results is also obtained for  higher Rydberg excitation $n$,
 albeit within a smaller range of $\delta$,
as presented in~\autoref{Fig:3b} 
for  the lowest-lying  bound state of the Rydberg molecular potential Rb($40S_{1/2}$)Rb($5S_{1/2}, F=1$).
 This is due to the smaller energy spacing between eigenvalues of the excited
 molecular potential for higher Rydberg excitations $n$. 
As a consequence, for large $\delta$,  the single-resonance approximation tends to fail and 
 $\alpha(\epsilon)$ should  be obtained using the coupled channel analysis. 
 For red detuning, $\Omega/(2\pi)=0.2$MHz\cite{units} and $\delta=10\gamma_m$, \autoref{Fig:3b} shows that 
 $\mathrm{Re}(\alpha$) can  be more than one order of magnitude larger than $\alpha_{\rm bg}$ within a timescale  that can be 
 conservatively estimated in the hundreds of microseconds range. 
 
Since our scheme is based on the virtual excitation of a single Rydberg molecular state as opposed to a pair of Rydberg atoms in the known schemes for atomic Rydberg dressing~\cite{Santos_2000, Johnson_2010,Pupillo_2010,Henkel_2010,Cinti_2010,Balewski_2014,Wang_2014,van_Bijnen_2015,Glaetzle_2015,Schempp_2015,Jau_2016,Aman_2016, Gaul_2016, Buchmann_2017, Khazali_2016,Li_2016}, the average density of the Rydberg-excitations in the ensemble can be kept lower. Consequently, collective decay effects~\cite{Zeiher_2016,Goldschmidt_2016} are here expected to be suppressed compared to the two-Rydberg-atoms dressing schemes.

%  Collective decay 
%  effects~\cite{Zeiher_2016,Goldschmidt_2016}  are here expected to be suppressed compared to known 
%  schemes for atomic Rydberg dressing~\cite{Santos_2000, Johnson_2010,Pupillo_2010,Henkel_2010,Cinti_2010,Balewski_2014,Wang_2014,van_Bijnen_2015,Glaetzle_2015,Schempp_2015,Jau_2016,Aman_2016, Gaul_2016, Buchmann_2017, Khazali_2016,Li_2016} due to  gas diluteness.

The thermally averaged scattering length is defined as 
\begin{align}
\bar{\alpha}(T)=\int\mathrm{d}\epsilon\alpha(\epsilon)P_T(\epsilon)\label{eqn:th_Av}\end{align}
where $\alpha(\epsilon)$ is determined from Eq.~\eqref{Eqn:scat}, and $P_T(\epsilon)=\exp{[-\epsilon/k_B T]}/\sqrt{\pi k_B\epsilon T}$ 
is the Maxwell-Boltzmann thermal distribution at temperature $T$, with $k_B$ being the Boltzmann-constant. 
\autoref{Fig:4} presents $\bar{\alpha}$ for the lowest lying potential of Rb($27S_{1/2}$)Rb($5S_{1/2}, F=1$) and two  
prototypical temperatures of  cold gases, $T=1$ and 5 $\mu$K\cite{units}. 
{ As this figure illustrates, the thermally averaged scattering length has a very similar detuning-dependence to the scattering length at a well-defined energy. However, due to the asymmetry of the scattering length with respect to energy, c.f. Eq~\eqref{Eqn:scat}, the averaged curves show an asymmetric character in the detuning $\delta$ as shown in the insets of \autoref{Fig:4}.}

Similar to \autoref{Fig:3a}, the ratio $\mathrm{Re}(\bar\alpha)/\mathrm{Im}(\bar\alpha)$ can easily reach $100$ or more, 
for a range of parameters that are within a reasonable experimental range.
This large ratio establishes the usefulness of Rydberg ORFs to manipulate interactions in a cold dilute atomic gas, 
which is one of the main results of this work.

%We propose that 
The results above can be generalised to a variety of different situations, including mixtures of two different alkali metal atoms,
having a negative scattering length  for the electron-atom collision~\cite{Greene_2000}.
Here, we consider  K and Cs as perturbing ground-state 
atoms with electron-atom scattering lengths provided in~\autoref{Tbl:perturber}. 

\begin{figure}[htp]

\subfloat[\label{Fig:5a}]{%
  \includegraphics[clip,width=\columnwidth]{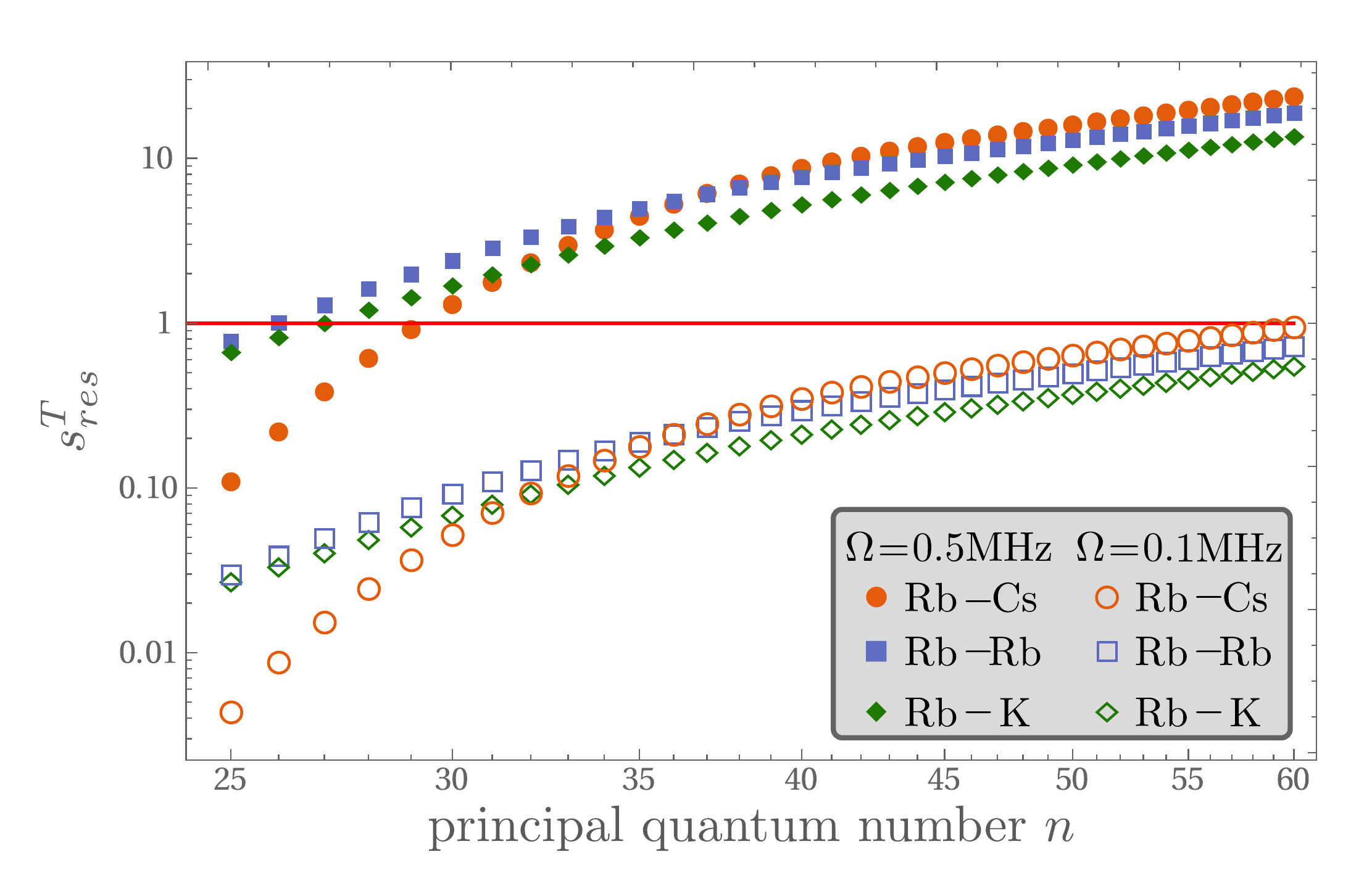}%
}

\subfloat[\label{Fig:5b}]{%
  \includegraphics[clip,width=\columnwidth]{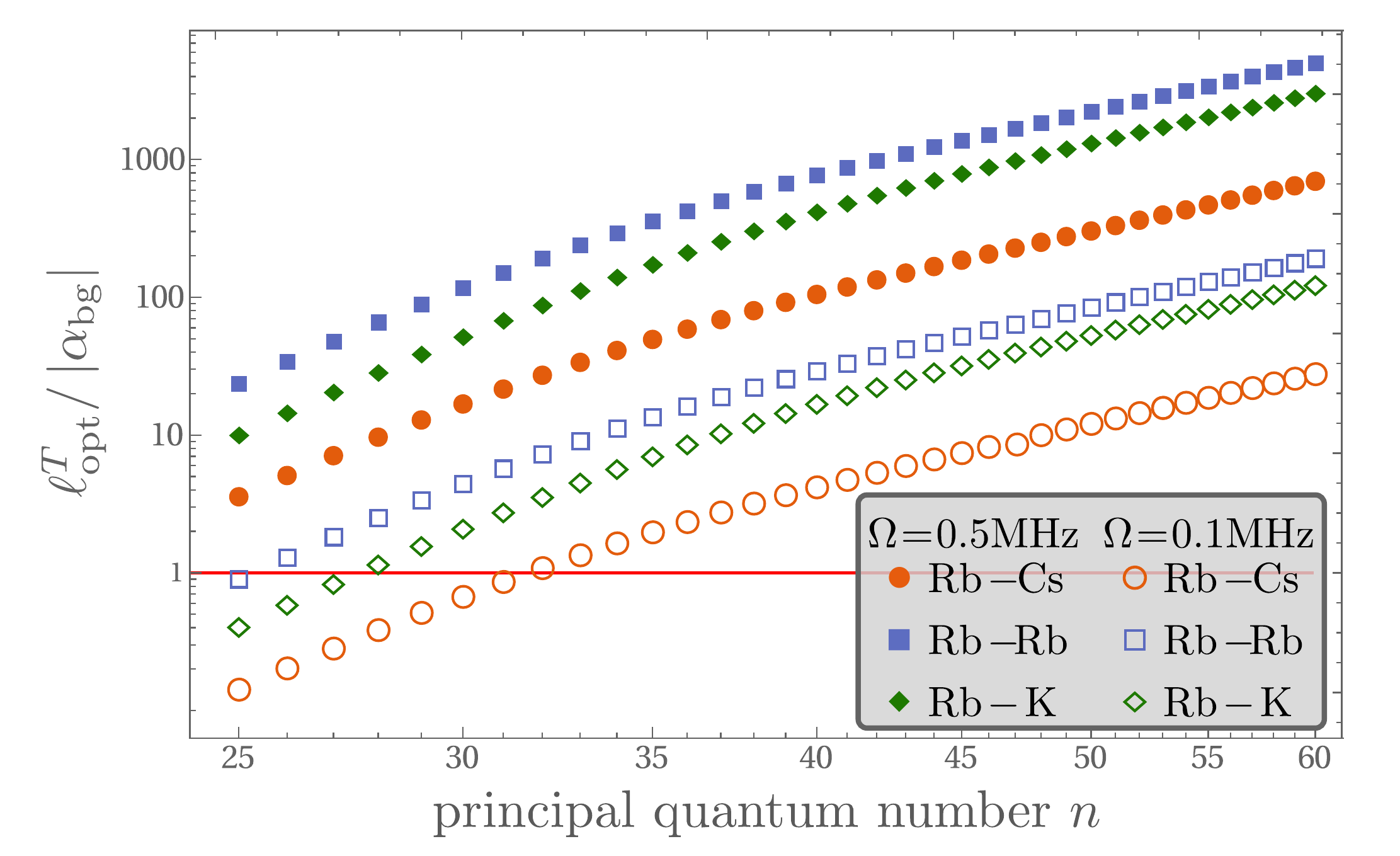}%
}

\caption{\label{Fig:5}For $T=1\mu K$, 
 (a) thermally averaged pole strength $s_{\rm res}^T$ and  (b)
  thermally averaged optical length $\ell_{\rm opt}^T$ 
  versus the Rydberg excitation $n$. See text and~\autoref{Tbl:perturber}.}
\end{figure}

% \begin{figure}[t]
%  \includegraphics[width=0.5\textwidth]{fig5a} \includegraphics[width=0.5\textwidth]{fig5b}
%  \caption{\label{Fig:5} For $T=1\mu K$, 
%  (a) thermally averaged pole strength $s_{\rm res}^T$ and  (b)
%   thermally averaged optical length $\ell_{\rm opt}^T$ 
%   versus the Rydberg excitation $n$. See text and Table~\ref{Tbl:perturber}. }
% \end{figure}

Further physical insight can be obtained by using an analytical approximation~\cite{Greene_2000} for the electronic 
Rydberg potential,
which is computed  considering only the Rydberg state Rb$(nS$) and the triplet  $s$-wave interaction between the Rydberg 
electron and the ground-state perturber.
In this approximation, the analytical Rydberg electronic potential is  
$V_U(R)=2\pi A_s^T[\kappa]\left|\psi_{ns}(R)\right|^2$,
with $\psi_{ns}(R)$ being  the Rydberg electron wave-function~\cite{functions}, and the bound state 
of its outermost lobe reproduces well the corresponding one of the exact Rydberg electronic 
potential~\cite{Greene_2000,Bendkowsky_2009, Bendkowsky_2010}, in which we are interested. 

In order to characterise the Rydberg OFR in an atomic ensemble at finite temperature, we introduce the average pole strength 
${s}^T_{\mathrm{res}}=\int_\mathrm{\epsilon}\mathrm{d}\epsilon\thinspace P_T(\epsilon) \ell_{\rm opt}(\epsilon)\gamma_m/(\bar{a}\bar{E})$ and optical length $ {\ell}^T_{\rm opt}={s}^T_{\mathrm{res}}\bar{a}\bar{E}/\gamma_m$,  $\ell_{\rm opt}(\epsilon)$ is the energy-dependent optical length defined in Eq.~\eqref{eqn:opt_length}. \autoref{Fig:5} shows that $\ell_{\rm opt}^T$ and $s_{\rm res}^T$ grow %algebraically with $n$ for large enough $n$ and, as expected, 
with the Rabi frequency as $\ell_{\rm opt}^T, s_{\rm res}^T\propto \Omega^2$, similar to usual OFRs. Remarkably, both quantities can be easily made much larger than one, $\ell_{\rm opt}^T/\bar{a}, s_{\rm res}^T \gg 1$,  by increasing the excitation of the Rydberg state $n$ or 
the Rabi frequency $\Omega$, within a reasonable experimental range. These results are in contrast
to, e.g.,  the measured OFRs in $^{88}$Sr, where $s_{\rm res}<1$ and are comparable to the best predicted values for $^{172}$Yb~\cite{Borkowski_2009}, however they can be obtained for all species, using reasonable experimental parameters.

\section{Summary}\label{sec:Sum}

In summary, we have proposed a novel mechanism for realising Feshbach resonances in cold gases using Rydberg molecular states. Since these Rydberg molecular states have long lifetimes and are present essentially for any atomic species, 
having a negative scattering length  for the electron-atom collision, we expect
that this technique can be directly applicable to a variety of situations where the given atoms do not enjoy magnetic Feshbach 
resonances. 
The present work opens up a host of new exciting directions. For example, while here we focus on $s$-wave collisions between alkali metals, in the future it will be interesting to explicitly address $p$-wave scattering. It will be exciting to extend this study
to alkaline-earth-type systems and atomic mixtures trapped in low-dimensional configurations, where even moderate optical tuning of the scattering length could significantly help the exploration of the strongly interacting regime~~\cite{Navon_2010,Kohstall_2012,Catani_2012,Koschorreck_2012}. %\cite{Navon_2010,Kohstall_2012,Catani_2012,Koschorreck_2012}. 
Coupling three-body and four-body resonances~\cite{Bendkowsky_2010}, which are well separated in the Rydberg molecular spectrum, could open the way to the realisation of effective multi particle interactions in cold gases.

\section*{Acknowledgements}
We thank I. I. Fabrikant for providing scattering lengths used in our calculations and J. Grimmel for his code to calculate the wave functions.  G.P. and N.S. are supported by the European Commission via ERC-St Grant ColdSIM (No. 307688). N. S. is subsequently supported via MSCA-IF-2015 grant (No. 706724). Additional partial support from OSMOZE (grant 35086UE), H2020 FET Proactive project RySQ (grant N. 640378),
ANR-FWF via "BLUESHIELD", and UdS via Labex NIE
and IdEX is gratefully acknowledged. R.G.F.  gratefully acknowledges financial support by the Spanish project 
FIS2014-54497-P (MINECO), and by the Andalusian research group FQM-207. 

\bibliographystyle{apsrev4-1}
\bibliography{literature}

\begin{thebibliography}{71}
\bibitem [{val()}]{values}%
  \BibitemOpen
  \href@noop {} {}\bibinfo {note} {In the present paper, we solve the ROFeR
  problem numerically for ${^{87}}Rb(5S_{1/2}, F=1)$ atoms. In the numerical
  coupled channel calculation the parameters $C_6=4422.673E_h a_0^6$,
  $C_8=550600 E_h a_0^8$ and $\sigma=5.90532 a_0$ are used in the Lennard-Jones
  potential in order to reproduce the scattering length 100.4 $a_0$ of
  ground-state atoms.}\BibitemShut {Stop}%
\end{thebibliography}

\end{document}